\documentclass[aps,superscriptaddress,nofootinbib,eqsecnum,prd,notitlepage,twocolumn]{revtex4-1} 

\pdfoutput=1

\usepackage{amsfonts}
\usepackage{amsmath}
\usepackage{amssymb}
\usepackage{graphicx,color}
\usepackage{float}
\usepackage{bbold}
\usepackage{mathrsfs}
\usepackage{hyperref}
\usepackage{subfigure}
\usepackage{dcolumn}
\usepackage{soul}
\usepackage{ulem}
\usepackage{verbatim}
\usepackage{bm}  
\usepackage{slashed}

\begin{document}

\title{Superconducting phase transition in planar fermionic  models with Dirac cone tilting}

\author{Y. M. P. Gomes}
\email{yurimullergomes@gmail.com}
\affiliation{Departamento de F\'{\i}sica Te\'{o}rica, Universidade do
  Estado do Rio de
Janeiro, 20550-013 Rio de Janeiro, RJ, Brazil}

\author{Rudnei O. Ramos}
\email{rudnei@uerj.br}
\affiliation{Departamento de F\'{\i}sica Te\'{o}rica, Universidade do
  Estado do Rio de
Janeiro, 20550-013 Rio de Janeiro, RJ, Brazil}
\affiliation{Physics Department, McGill University, Montreal, Quebec, H3A 2T8, Canada}

\begin{abstract}

The chiral and superconducting gaps are studied in the context of a
planar fermion model with four-fermion interactions. The effect of the
tilt of the Dirac cone on both gaps is shown and discussed. Our
results point to two different behaviors exhibited by planar fermionic
systems.  We show that there is a threshold value $\tilde{t}^*$ for
the effective tilt parameter such that when $|{\bf \tilde{t}}| <
\tilde{t}^*$,  the superconducting phase persists for negative values
of the superconducting coupling constant. {}For positive values of the
superconducting coupling constant, the induction of a superconducting
gap by a chemical potential exists and which is similar to the one
seen in graphene-like systems.  {}For  $|{\bf \tilde{t}}| >
\tilde{t}^*$ and a negative superconducting coupling constant, the
superconducting phase can be present, but it is restricted to a
smaller area in the phase portrait. Our analysis also shows that 
when $|{\bf \tilde{t}}| > \tilde{t}^*$ and for positive values for 
the superconducting coupling constant, the induction of a superconducting 
gap in the presence of a chemical potential is ruled out. In this case, 
the increase of the chemical potential works in favor of the manifestation 
of a metallic phase.

\end{abstract}

\maketitle
\section{Introduction}
\label{sec:intro}

The possibility of the superconducting phase in the Weyl fermion
system is one of the popular topics in condensed matter physics. The
discovery of a tilted Weyl dispersion in realistic materials, for
example, type-II Weyl semimetals, has accelerated the related research
on this topic.  The detailed verification of the phase diagram for
these types of systems is of relevance for researchers in the field.
Here we approach this important problem from the point of view of
quantum field theory techniques.  Since the seminal work of Gross and
Neveu~\cite{intro1}, where the authors use quantum field theory (QFT)
tools to describe two-dimensional massless fermions with quartic
interactions, much attention was expended to apply QFT techniques in
low-dimensional systems and with special attention to condensed matter
problems.  One of the most interesting examples of the applications of
QFT in condensate matter is the study of graphene~\cite{intro2}. In
this almost planar system, the electrons obey linearly dispersing
relations and the fermionic excitations are well described by a
relativistic Dirac equation in (2+1)-dimensions.      

The Lorentz symmetry is respected by the electrons in graphene due to
its relativistic characteristics, but this feature is an exception
compared to the majority of materials in condensed matter.  Some of
the condensed matter systems, where the dispersion in the proximity of
band touching points can be generically linear and resemble the Weyl
equation, do not respect Lorentz
symmetry~\cite{Grassano2020,zhang2021,kost2022,Tamashevich2022}.  Even
though there are quasiparticles in the aforementioned systems that
behave like Weyl fermions~\cite{weyl1}, these systems are described by
Weyl-like Hamiltonians and, thus, these quasiparticles are by
construction massless and more stable against gap formation in
comparison to Dirac ones~\cite{weyl2}.

Our proposal in this paper is to study how the properties associated
with Weyl fermions influence the formations of chiral and
superconducting gaps in planar systems. Superconductivity was studied
in three-dimensional Weyl semi-metals of both types I and II with a
particular effect of the tilting of the Dirac
cone~\cite{Rosenstein2018}.  {}From the experimental side, despite the
challenges, there have been studies in heterostructures consisting of
thin films of half-metal and spin-singlet
superconductor~\cite{hao2017}. The theoretical study of
superconducting instabilities in Dirac and Luttinger fermions has also
been recently analyzed~\cite{szabo2021}. Here we will extend the usual
Weyl Hamiltonian used in the description of Weyl semimetals
(WSM)~\cite{goerbig1,goerbig2} by introducing two forms of
four-fermion interactions that will allow for a chiral phase and a
superconducting phase. We also analyze the properties of this system
under the effects of a finite chemical potential, which in practice
models the doping process.  This will allow us to study the allowed
phase transitions in a $(2 + 1)$- dimensional Gross-Neveu (GN)-type model, 
which
describes the competition between the chiral symmetry breaking and
superconductivity. These two phenomena will dispute the true ground
state of the system through the intensity of the coupling constants
and as a function of the chemical potential. Let us also recall that
chiral symmetry and its breaking can be seen as a way to describe
the metal-insulator phase transition in these planar systems. Thus,
the study of chiral symmetry breaking in planar systems by GN-like
four-fermion interactions has become a useful tool for qualitative
analysis of the two-dimensional system and has already been used
successfully in many different
contexts~\cite{Caldas:2008zz,Caldas:2009zz,gn1,gn2,gn3,Ramos:2013aia,Klimenko:2012tk,Klimenko:2013gua,Ebert:2015hva,Ebert:2016ygm,Zhukovsky:2017hzo,Zerf:2017zqi,Fernandez:2021dfk,Drut:2007zx,Juricic:2009px,Herbut:2009vu,Rostami:2020set,Khunjua:2021hhb,Khunjua:2021fus}.

In this paper, we also want the address the question of the production
of a superconducting phase in the model and how the tilting of the
Dirac cone affects it. The phenomenon of electron pairing in the vast
majority of superconductors follows the Bardeen–Cooper–Schrieffer
(BCS) theory of superconductivity. The BCS theory describes the
condensation of electrons into pairs with anti-parallel spins in a
singlet state with an $s$-wave symmetry. The $s$-wave channel will be
the superconducting channel that will be addressed in this
paper. Several works have already indicated that superconductivity
appears in planar systems, such as twisted bilayer
graphene~\cite{superc1}, normal trilayer graphene~\cite{superc2}, and
twisted trilayer graphene as well~\cite{superc3,superc4}. However, here we address the the effects
caused by the tilt of the Dirac cone on the combined chiral and
superconducting phases and how it might influence, in particular, the
superconducting gap. 

The tilting, the coupling constants for the chiral and pairing
interactions in the superconducting channel, and the chemical
potential provide four independent parameters. {}From the coupling
constants, we can present a phase diagram in the case where the tilt
factor and the chemical potential take values that are of practical
interest. In particular, we find  that the existence
of the superconducting phase strongly depends on whether the tilt
factor is larger or smaller than a threshold value, $\tilde{t}^*$,
which we explicitly estimate both analytically and numerically,
besides of depending as well on the sign of the pairing interaction in
the superconducting channel.

The remainder of this paper is organized as follows. In
Sec.~\ref{sec2}, we briefly discuss the main properties of
two-dimensional Dirac and Weyl semimetal systems. In Sec.~\ref{sec3},
we present the extension of the model that describes the four-fermion
interactions for the excitonic and superconducting channels. The
effective thermodynamic potential for the system is derived through
the mean-field and one-loop semi-classical approximation level. The
effects of he anisotropy, tilting of the Dirac cone, and chemical
potential are taking into account in this derivation. In
Sec.~\ref{sec4}, we show and discuss the effect of the chemical
potential $\mu$ on the effective thermodynamic potential and we
present the chiral and superconductivity gap equations of the
system. In Sec.~\ref{sec5}, we discuss the phase transition of the
system as a function of the chemical potential. {}In
Sec.~\ref{conclusions}, our conclusions and remarks are presented,
along also with the discussion of the possible implications of our
results to some current experimental planar materials of interest. Two
appendices are also included where some technical details are
presented.  Throughout this paper, we will be considering the natural
units where $\hbar = k_B = c =1$.

\section{Two dimensional Weyl semimetals}
\label{sec2}

In this section, we present the main details of the representation of
the low energy electronic excitations in the two-dimensional Weyl
semimetals. Within the tight-binding approximation calculated for the
honeycomb-like lattices,  the low energy dynamics of the
two-dimensional system of Weyl fermions can be described by the
Hamiltonian~\cite{goerbig1,goerbig2}
\begin{equation}\label{eqenergy}
H_t({\bf p}) = v_F \left[ ( {\bf t} \cdot {\bf p} )\tau^0 + (\xi_x
  p_x) \tau^x + (\xi_y p_y) \tau^y \right],
\end{equation}
where $v_F$ is the {}Fermi velocity, ${\bf t}$ is called the tilt
vector and that describes the Dirac cone tilt, ${\bf \xi}=(\xi_x,
\xi_y)$ is the vector that describes the anisotropy of the material,
$\tau^0 = \mathbb{1}$ is the $2 \times 2$ identity matrix and
$\tau^{x,y}$ are the Pauli matrices. In the limit ${\bf t} \rightarrow
0$ and $\xi_x=\xi_y=1$, we recover the Hamiltonian of the isotropic
graphene.  The tilt vector ${\bf t}$ is related to the separation
between the Dirac cones in the Weyl semi-metal. A consequence of the
non-null tilt term in Eq.~\eqref{eqenergy} is that the Dirac points,
denoted by $D$ and $D'$, no longer coincide with the Brillouin corners
K and $K'$ (see, e.g.,  Ref.~\cite{goerbig1}). In particular, type-I
Weyl semi-metals are characterized by $|{\bf t}|<1$, while type-II
ones are characterized  by $|{\bf t}|>1$. {}From the Hamiltonian given by
Eq.~\eqref{eqenergy}, one finds that the spectrum is given by
\begin{equation}
E_\lambda({\bf p}) = v_F \Big[{\bf t} \cdot {\bf p} + \lambda
  \sqrt{(\xi_x p_x)^2 + (\xi_y p_y)^2} \Big],
\end{equation}
where $\lambda = \pm 1$ represent the conduction and valence bands,
respectively. Note that to be able to associate $\lambda=+1$ with a
positive and $\lambda=-1$ with a  negative energy state, it is
required that~\cite{goerbig1,goerbig2}
\begin{equation}\label{cond}
\sqrt{\left( \frac{t_x}{\xi_x}\right) ^2+
  \left(\frac{t_y}{\xi_y}\right)^2} = |{\bf \tilde{t}}| <1,
\end{equation}
where $|{\bf \tilde{t}}|$ is called the {\it effective tilt
  parameter}. 

The Hamiltonian given by Eq.~\eqref{eqenergy} commutes with the chirality
operator defined as
\begin{equation}
\mathcal{C} = \frac{(\xi_x p_x)\tau_x+(\xi_y p_y)\tau_y}{\sqrt{(\xi_x
    p_x)^2 + (\xi_y p_y)^2}},
\end{equation}
with the eigenvalues given by $\alpha= \pm 1$.  Taking into account
all the degeneracies of the system, the free Weyl fermion can be
described with a four-component spinor and a Dirac-like Lagrangian
density can be written as follows (see also, e.g.,
Ref.~\cite{Gomes:2021nem}):
\begin{eqnarray}
\mathcal{L} &=& \sum_{k=1}^N i \bar{\psi}_k M^{\mu \nu}\gamma_\mu
\partial_\nu\psi_k ,
\label{Lagr}
\end{eqnarray}
where $\psi$ is a four-component Dirac fermion. The $\gamma$-matrices
are written as
\begin{equation}
\gamma^\mu = \tau^\mu \otimes \begin{pmatrix} 1 & 0 \\ 0 & -1
\end{pmatrix},
\end{equation}
with $\mu = 0,1,2$, $\tau^\mu = (\tau_z, i \tau_x , i \tau_y)$,
$\bar{\psi} = \psi^\dagger \gamma^0$ and $\tau_z$ is the third Pauli
matrix. The $\gamma$-matrices obey the identity $\gamma^\mu \gamma^\nu
= \eta^{\mu \nu} + i\epsilon^{\mu \nu \lambda} \gamma_3
\gamma_\lambda$, where $\gamma_3 = \begin{pmatrix} \mathbb{1} & 0\\ 0
  & -\mathbb{1}
\end{pmatrix}$ and $\eta^{\mu \nu}= diag (+,-,-)$. Thus, it is straightforward to prove that the $\gamma$-matrices obey the algebra $\{\gamma^\mu, \gamma^\nu \}= 2 \eta^{\mu \nu}$. The matrix $M$ in Eq.~(\ref{Lagr}) is explicitly given by
\begin{equation} \label{Mmatrix}
M = \begin{pmatrix} 1 & - v_F t_x & -v_F t_y\\ 0 & -v_F \xi_x & 0\\ 0
  & 0 & -v_F\xi_y 
\end{pmatrix}.
\end{equation}
We can see $M$ as representing an analogous of an effective
metric. One also notices that $M^{\mu \nu}$ contains the parameters
that explicitly break the Lorentz symmetry, which is a consequence of
the tilting of the Dirac cone. It is easy to show that the Lagrangian
density given by Eq.~(\ref{Lagr}) has a discrete chiral symmetry given by $\psi
\rightarrow \gamma_5 \psi$ and $\bar{\psi} \rightarrow -\bar{\psi}
\gamma_5$, with
\begin{equation}
i \gamma_5 = \begin{pmatrix} 0 & \mathbb{1}\\ -\mathbb{1} & 0
\end{pmatrix}.
\end{equation}
Throughout the next sections, one follows the
Ref.~\cite{Gomes:2021nem} and choose the mass term that breaks the
chiral symmetry as $\bar{\psi} \psi$. 

\section{Chiral and difermion interactions}
\label{sec3}

To write an effective Lagrangian density that can describe the
(2+1)-dimensional Weyl semimetal with both chiral symmetry breaking
(excitonic pairing) and superconductivity (Cooper pairing), two forms
of four-fermion interactions can be
introduced~\cite{Klimenko:2012tk}. One of them is a four fermion
interaction for the scalar fermion-antifermion and the other one is
for the scalar difermion channel. The complete model can then be
written as
\begin{eqnarray}
\mathcal{L} &=& \sum_{k=1}^N \bar{\psi}_k( i M^{\mu \nu}\gamma_\mu
\partial_\nu + \gamma^0 \mu)\psi_k + \frac{G_1
  v_F}{2N}\Bigg(\sum_{k=1}^N \bar{\psi}_k \psi_k \Bigg)^2 \nonumber
\\ &&+ \frac{G_2 v_F}{2N}\sum_{k=1}^N (\psi_k^T C
\psi_k)\sum_{j=1}^N(\bar{\psi}_j C \bar{\psi}_j^T),
\end{eqnarray}
where  $C = i \gamma^2$ is the charge conjugation matrix and $G_1$ and
$G_2$ are the coupling constants for the chiral and difermion
channels. The coupling constants $G_1$ and $G_2$ are negative for an
attractive interaction, while they are positive for a repulsive
interaction. The attractive/repulsive nature of the couplings will be
decisive for the phase transition patterns analyzed in the subsequent
sections. The effective action of the model can be expressed as
\begin{eqnarray}
\exp( i S_{eff}) &=&\int D\bar{\psi} D\psi D \Delta  D \Delta^* D
\sigma  \nonumber \\ &\times & \exp \left\{ \int d^3x \left[
  \frac{N}{2 G_1 v_F} \sigma^2 + \frac{N}{2 G_2 v_F}\Delta^* \Delta
  \right. \right.  \nonumber \\ &+ &  \left. \left. \sum_{k=1}^N
  \bar{\psi}_k(i M^{\mu \nu}\gamma_\mu  \partial_\nu  + \gamma^0 \mu+
  \sigma )\psi_k \right. \right.  \nonumber \\ &+ &
  \left. \left. \frac{\Delta^*}{2} \psi_k^T C \psi_k +
  \frac{\Delta}{2} \bar{\psi}_k C \bar{\psi}_k^T \right] \right\} ,
\label{effaction}
\end{eqnarray}
where $\sigma = \frac{G_1 v_F}{N} \sum_{j=1}^N \bar{\psi}_j \psi_j$,
$\Delta =  \frac{G_2 v_F}{N}\sum_{j=1}^N \psi_j^T C \psi_j$ and
$\Delta^* = \frac{G_2 v_F}{N}\sum_{j=1}^N \bar{\psi}_j C
\bar{\psi}_j^T$. We can explicitly integrate over the fermion field
(for the technical details, see Appendix~\ref{appA}) and the effective
action can be rewritten as $S_{eff}(\sigma, \Delta, \Delta^*) = N \int
d^3x \Omega(\sigma, \Delta, \Delta^*)$, where $\Omega$ is the
effective thermodynamics potential,
\begin{eqnarray}
\Omega(\sigma, \Delta, \Delta^*) &=&   \frac{1}{2 G_1 v_F}
\sigma^2 + \frac{1}{2 G_2 v_F}\Delta^* \Delta   \nonumber \\ &+&
\sum_{i=1}^2 \int \frac{d^3p}{(2 \pi)^3} ln \lambda_i(p),
\end{eqnarray}
with $\lambda_i$ denoting the eigenvalues of $B = C D C^{-1} D^T -
|\Delta|^2$, with $D = M^{\mu \nu}\gamma_\mu  \partial_\nu + \gamma^0
\mu- \sigma$, which are given by 
\begin{eqnarray}
\lambda_{1,2} &=& \sigma^2 + \left[p_0 - v_F({\bf t} \cdot {\bf
    p})\right]^2 - v_F^2{\bf \tilde{p}}^2 - \mu^2 - |\Delta|^2
\nonumber \\ && \pm 2 \sqrt{\sigma^2\left\{\left[p_0 - v_F({\bf t}
    \cdot {\bf p})\right]^2 - v_F^2{\bf \tilde{p}}^2\right\} +
  v_F^2\mu^2{\bf \tilde{p}}^2 }.  \nonumber \\
\end{eqnarray}
with ${\bf \tilde{p}} = (\xi_x p_x, \xi_y p_y)$. Using the identity
\begin{eqnarray}
\int_{-\infty}^\infty dp_0 \ln(p_0 - A) = i \pi |A| ~,
\end{eqnarray}
we find that
\begin{eqnarray}
 \sum_{i=1}^2 \int \frac{d^3p}{(2 \pi)^3} \ln \lambda_i(p) &=& - \int
 \frac{d^2p}{(2 \pi)^2} (|\Sigma^+| + |\Sigma^-|), \nonumber \\
\end{eqnarray}
where 
\begin{eqnarray}
\Sigma^\pm = v_F({\bf t} \cdot {\bf p}) +\sqrt{ \tilde{E}^2 + \mu^2 +
  |\Delta|^2 \pm 2 \sqrt{\sigma^2 |\Delta|^2 + \mu^2 \tilde{E}^2 } },
\nonumber\\
\end{eqnarray}
and $\tilde{E}^2 = v_F^2{\bf \tilde{p}}^2 + \sigma^2$.  {}Finally, for
constant configurations $\sigma_0 = \langle \sigma \rangle$ and
$\Delta_0 = \langle \Delta \rangle = \langle \Delta^* \rangle$, we
find
\begin{eqnarray}
\Omega(\sigma_0,\Delta_0, \mu) &=&  \frac{1}{2 G_1 v_F}
\sigma_0^2 + \frac{1}{2 G_2 v_F}\Delta_0^2  \nonumber \\ &-& \int
\frac{d^2p}{(2 \pi)^2} \left( |\Sigma_0^+| + |\Sigma_0^-|  \right),
\label{effpotbare}
\end{eqnarray}
where $\Sigma^\pm_0 = \Sigma^\pm (\sigma=\sigma_0,
\Delta=\Delta_0)$. Note that the momentum integral in
Eq.~(\ref{effpotbare}) is  divergent in the ultraviolet limit and, thus, the
effective potential given by Eq.~(\ref{effpotbare}) needs to be
renormalized. The renormalization of Eq.~(\ref{effpotbare}) is
described below.

\subsection{Renormalization}

Taking $\mu=0$ in Eq.~(\ref{effpotbare}), we will have that
$\Sigma^\pm_0   = {\bf t} \cdot {\bf p} + \sqrt{ {\bf \tilde{p}}^2 +
  (\sigma_0 \pm \Delta_0)^2} $ and, therefore,
\begin{eqnarray}
\Omega(\sigma_0, \Delta_0) &=&  \frac{1}{2 G_1 v_F} \sigma_0^2 +
\frac{1}{2 G_2 v_F} \Delta_0^2  \nonumber\\ &-& \int
\frac{d^2p}{(2 \pi)^2}\left| v_F({\bf t} \cdot {\bf p}) + \sqrt{
  v_F^2{\bf \tilde{p}}^2 + (\sigma_0 + \Delta_0)^2}\right|  \nonumber
\\ &- & \int \frac{d^2p}{(2 \pi)^2} \left|v_F({\bf t} \cdot {\bf p}) +\sqrt{
  v_F^2{\bf \tilde{p}}^2 + (\sigma_0 - \Delta_0)^2} \right|.  \nonumber
\\
\label{unrenor}
\end{eqnarray}
The linear term ${\bf t} \cdot {\bf p}$ in Eq.~(\ref{unrenor})
vanishes in the integration over the angular variable\footnote{We use
  the identity $\int_0^{2 \pi} d \theta | a \cos \theta + b| = 2 \pi b
  \Theta(b-a)$, where $ \Theta(x)$ is the Heaviside function, for
  $a>0$ and $b>0$.}, but the integral in Eq.~\eqref{unrenor} is still
divergent. Thus, applying the re-scaling  $\xi_{x,y}p_{x,y}
\rightarrow p_{x,y}$ and integrating with the introduction of a
momentum cutoff $\Lambda$, one defines the renormalization conditions,
\begin{eqnarray}
\frac{1}{g_1(m)} &=&  v_F\frac{d^2\Omega(\sigma_0,\Delta_0)}{d
  \sigma_0^2}\Big{|}_{\sigma_0=m,\Delta_0=0}
\nonumber\\ &=&\frac{1}{G_1} + \frac{2 m}{\pi v_F\xi_x \xi_y}-
\frac{\Lambda}{\pi v_F\xi_x \xi_y},
\end{eqnarray}
and 
\begin{eqnarray}
\frac{1}{g_2(m')} &=&  v_F \frac{d^2\Omega(\sigma_0,\Delta_0)}{d
  \Delta_0^2}\Big{|}_{\sigma_0=0,\Delta_0=m'}
\nonumber\\ &=&\frac{1}{G_2} + \frac{2 m'}{\pi v_F \xi_x \xi_y}-
\frac{\Lambda}{\pi v_F\xi_x \xi_y},
\end{eqnarray}
where $m$ and $m'$ are regularization scales. Going further, defining
the renormalized couplings $g_1$ and $g_2$ as
\begin{eqnarray}
\frac{1}{g_1} = \frac{1}{g_1(m)}-\frac{2 m}{\pi  v_F \xi_x \xi_y},
\end{eqnarray}
and 
\begin{eqnarray}
\frac{1}{g_2} = \frac{1}{g_2(m')}-\frac{2 m'}{\pi v_F \xi_x \xi_y},
\end{eqnarray}
the renormalized effective thermodynamic potential finally can be
expressed as
\begin{eqnarray}
\Omega^{\text{ren}}(\sigma_0, \Delta_0) &=&  \frac{1}{2 g_1 v_F}
\sigma_0^2 + \frac{1}{2 g_2 v_F} \Delta_0^2   \nonumber
\\  &&  \hspace{-1.cm} + \frac{(\sigma_0+\Delta_0)^3}{6 \pi v_F^2
  \xi_x \xi_y}  + \frac{|\sigma_0-\Delta_0|^3}{6 \pi v_F^2 \xi_x
  \xi_y}  ~. 
\label{effpot}
\end{eqnarray}

\subsection{Phase diagram of the system at $\mu=0$}

Let us first specialize in the analysis of the effective thermodynamic
potential and its properties in the case of a null chemical
potential. In this perspective, we analyze the two sectors, the chiral
and the superconductor ones, individually. This will allows us to
extract the main characteristics of the model. After this analysis, we
can then compare the results and show where each phase will be
mandatory in the system.  

The minima of Eq.~(\ref{effpot}) are given in terms of the gaps
$\bar{\sigma}_c$ and $\bar{\Delta}_c$, which are defined as
$\bar{\sigma}_c = \pi v_F\xi_x \xi_y/|g_1|$ and $\bar{\Delta}_c = \pi
v_F \xi_x \xi_y/|g_2|$.  By analyzing the thermodynamic potential given by
Eq.~(\ref{effpot}), in the absence of tilting, it can be established
that the system can be characterized by three phases, according to the
values of $\sigma$, $\Delta$, and the coupling constants. We follow
the same classification used in Ref.~\cite{Klimenko:2013gua} which
studied the nontilted system. {\it Phase I}: this is the symmetric
phase, where both vacuum expectations values for the chiral and
superconducting phases are zero, $\sigma_0 = \Delta_0 = 0$, and which
can take place when for $g_1>0$ and $g_2>0$. {\it Phase II}: in this
phase  $\sigma_0 = \bar{\sigma}_c \neq 0$ and  $\Delta_0 =0$ and it
can happen when $g_1 <0$. {\it Phase III}: in this phase $\sigma_0=0$
and $\Delta_0 =\bar{\Delta}_c  \neq 0$ and it can happen when $g_2<0$.
When $g_1$ and $g_2$ are simultaneously negative, the system is
characterized as phase II if $|g_1|>|g_2|$ and as
phase III for $|g_1|<|g_2|$.  In the next two sections, we will
analyze how the effects of both tilting and chemical potential affect
these different phases allowed by the model.

\section{Tilting effects on the superconducting gap}
\label{sec4}
 
Let us now turn on the effects of the tilting of the Dirac cone on
the different three phases allowed by the model and described at the
end of Sec.~\ref{sec3}. It is useful to first focus on the
pure chiral phase (when $\Delta_0 =0$), where we here briefly
reproduce some of the results obtained in Ref.~\cite{Gomes:2021nem}.
After that, we will analyze the case of the
superconducting gap in details.

\subsection{The pure chiral phase ($\Delta_0 =0$)}

By considering the pure chiral phase, i.e., by considering $\Delta_0
=0$, one notices that
\begin{equation}
(\Sigma^\pm_0)|_{\Delta_0=0} = \mathcal{E}^\pm_\sigma =v_F({\bf t}
  \cdot {\bf p}) +\big| \tilde{E}_0 \pm \mu \big| ,
\end{equation}
where $\tilde{E}_0 = \sqrt{v_F^2 {\bf \tilde{p}}^2 +
  \sigma_0^2}$. Assuming $\mu>0$, one finds in this case that the
effective thermodynamic potential~\eqref{effpot} becomes
\begin{eqnarray}
\lefteqn{\Omega^{ren}(\sigma_0,0, \mu ) =\frac{\sigma_0^2}{2 g_1 v_F}
  + \frac{\sigma_0^3}{3 \pi v_F^2 \xi_x \xi_y} } \nonumber \\ &&- \int
\frac{d^2p}{(2 \pi)^2}\Bigg(|\mathcal{E}^+_\sigma| +
|\mathcal{E}^-_\sigma| - 2 E_\sigma \Big) \nonumber \\ &&=
\frac{\sigma_0^2}{2 g_1 v_F} + \frac{\sigma_0^3}{3 \pi v_F^2 \xi_x
  \xi_y} - \int \frac{d^2p}{(2 \pi)^2}\Big(\mu- \tilde{E}_\sigma +
|\mu-\tilde{E}_\sigma | \Big) \nonumber \\ &&= \frac{\sigma_0^2}{2 g_1
  v_F} + \frac{\sigma_0^3}{3 \pi v_F^2 \xi_x \xi_y} -2  \int
\frac{d^2p}{(2 \pi)^2}(\mu - \tilde{E}_\sigma) \Theta(\mu-
\tilde{E}_\sigma),  \nonumber \\
\label{potchiral}
\end{eqnarray}
with $\tilde{E}_\sigma =v_F({\bf t} \cdot {\bf p}) + \sqrt{v_F^2{\bf
    \tilde{p}}^2+ \sigma_0^2}$ and we have used the identity $x + |x|
= 2 x \Theta(x)$.  {}From Eq.~\eqref{potchiral}, one can derive the
gap equation,
\begin{equation}\label{gapchiral}
1 + \frac{{\rm sign}(g_1)\sigma_0}{\bar{\sigma}_c} +2 g_1 \int
\frac{d^2p}{(2 \pi)^2} \frac{\Theta(\mu-
  \tilde{E}_\sigma)}{\sqrt{v_F^2{\bf \tilde{p}}^2+ \sigma_0^2}} =0.
\end{equation}
It follows from Eq.~(\ref{gapchiral}) that for $g_1>0$ the chiral
symmetry is maintained for any $\mu>0$.  We can also see that the
effect of the effective tilt parameter $|{\bf \tilde{t}} |$ in
Eq.~(\ref{gapchiral}) is to enhance the effect of the chemical
potential and, hence, to lower the point of chiral symmetry
restoration. In particular, for $g_1<0$ the chiral symmetry breaks for
$\mu< \mu_c$ and is restored for $\mu >\mu_c$, where the critical
chemical potential $\mu_c$ is found to be given
by~\cite{Gomes:2021nem}
\begin{equation}
\mu_c= \sqrt{1-|{\bf \tilde{t}}|^2} \bar{\sigma}_c.
\end{equation}
Therefore, we can say that the presence of the non-vanishing tilt
parameter tends to facilitate the chiral symmetry restoration.  One
also finds from  Eq.~(\ref{gapchiral}) that the chiral order
parameter, which is the solution of \eqref{gapchiral}, jumps
discontinuously from $\sigma_0=\bar{\sigma}_c $ to $\sigma_0 = 0$ as
we change the chemical potential from $\mu<\mu_c$ to $\mu>\mu_c$. This
is a first-order transition that exists for both the non-tilted
case $|{\bf \tilde{t}}|=0$ and the tilted case $|{\bf \tilde{t}}|
\neq 0$. 

The charge density $n$ is defined as
\begin{equation}
n = - N\frac{\partial \Omega^{ren}(\sigma_0,0,\mu)}{\partial \mu}
\Bigg{|}_{\sigma_0 = \langle \sigma_0 \rangle}.
\end{equation}
The exact expression for $n$ can be readily calculated from
$\Omega^{ren}(\sigma_0,0,\mu)$ and it reads
\begin{eqnarray}
n(g_1>0) = \frac{N\mu^2}{2 \pi v_F^2 \xi_x \xi_y(1-|{\bf
    \tilde{t}}|^2)^{3/2}} ,
\end{eqnarray}
when $g_1>0$ and
\begin{equation}
n(g_1<0 ) = \frac{N(\mu^2 - \mu_c^2)}{2\pi v_F^2 \xi_x \xi_y(1-|{\bf
    \tilde{t}}|^2)^{3/2}}\Theta \left(\mu^2 -\mu_c^2 \right),
\end{equation}
when $g_1< 0$.

In the next section one turns to the analysis of the
superconducting phase. 

\subsection{The pure superconducting phase ($\sigma_0=0$)}

\begin{center}
\begin{figure}[!htb]
\subfigure[$g_2<0$]{\includegraphics[width=7.5cm]{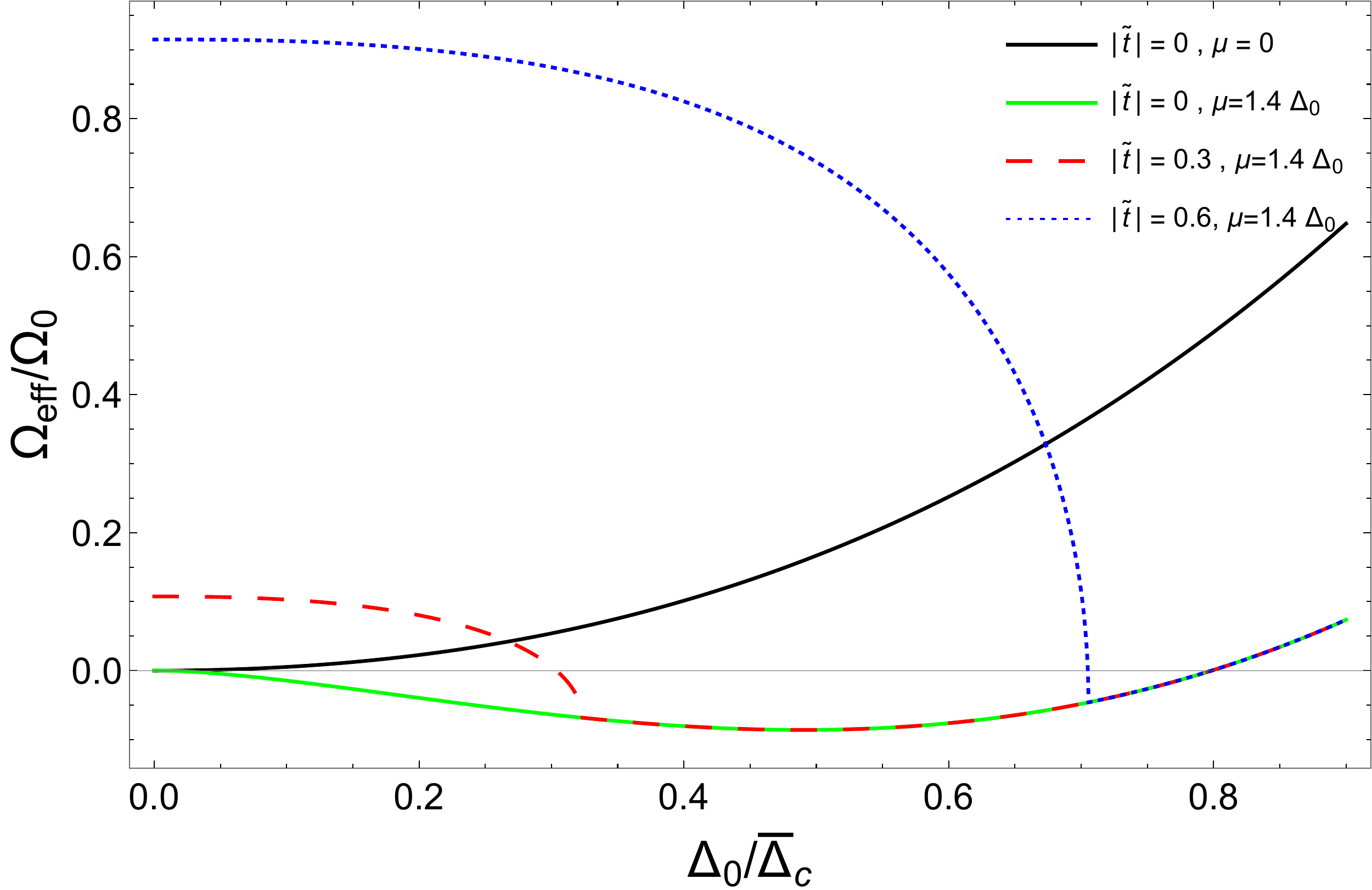}}
\subfigure[$g_2>0$]{\includegraphics[width=7.5cm]{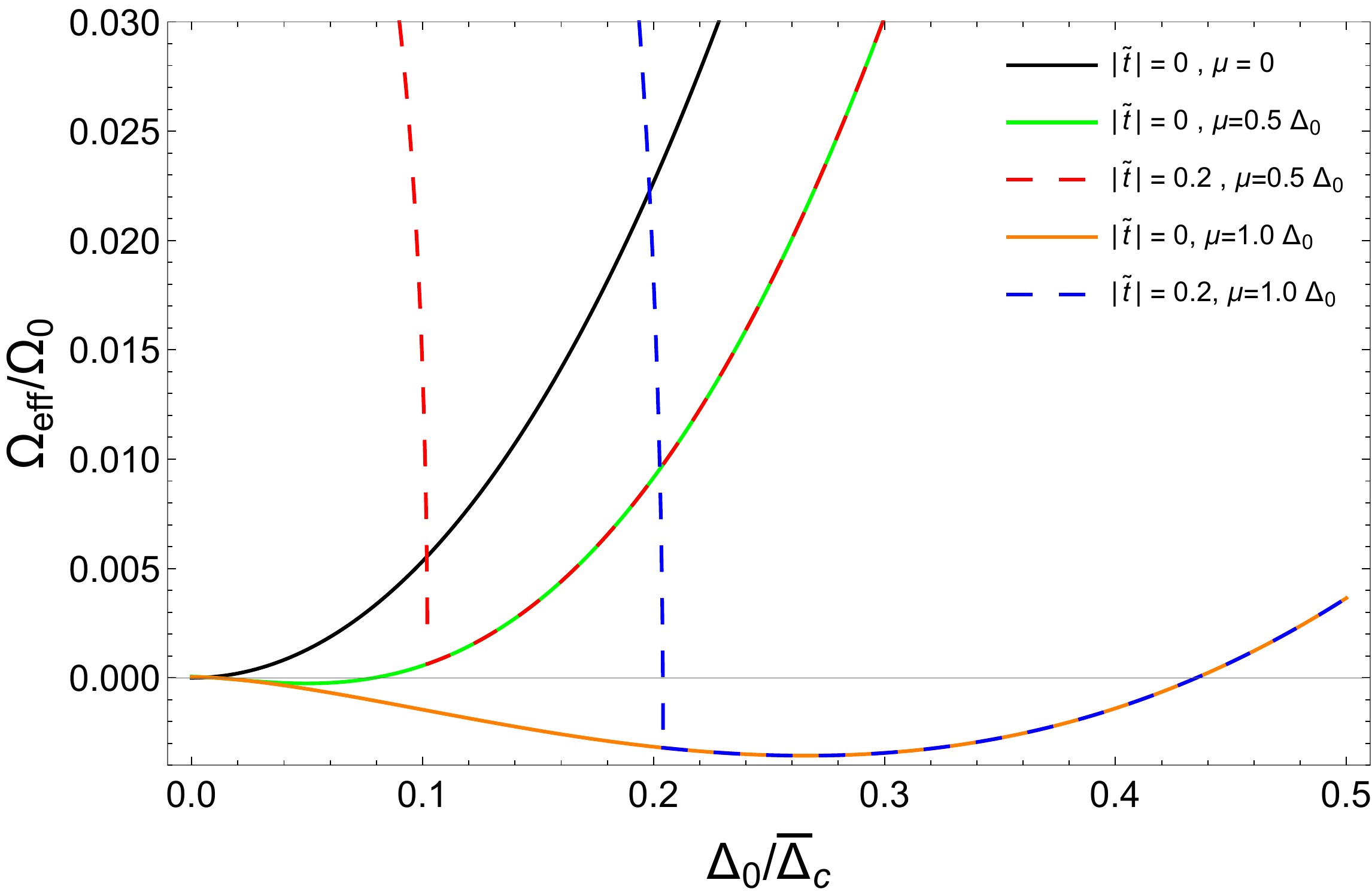}}
\caption{Effective thermodynamic potential for (a) $g_2<0$  and
(b) $g_2>0$, in units of $\Omega_0 = N
  \bar{\Delta}_c^3/(\pi v_F^2 \xi_x \xi_y)$ as a function of
  $\Delta/\bar{\Delta}_c$.}
\label{fig1}
\end{figure}
\end{center}

In the case of a pure superconducting phase, i.e., considering now
$\sigma_0=0$, and using the identity 
\begin{equation}
(\Sigma^\pm_0)|_{\sigma_0=0}  = \mathcal{E}^\pm_\Delta =v_F({\bf t}
  \cdot {\bf p}) +\sqrt{ (v_F|{\bf \tilde{p}}| \pm \mu)^2 + \Delta_0^2
  } ,
\end{equation}
the effective thermodynamic potential~\eqref{effpot} can be written as
\begin{widetext}
\begin{eqnarray}
 \Omega^{ren}(0,\Delta_0, \mu ) &=& \frac{\Delta_0^2}{2 g_2 v_F} +
 \frac{\Delta_0^3}{3 \pi v_F^2 \xi_x \xi_y}  - \int \frac{d^2p}{(2
   \pi)^2}\left\{ |\mathcal{E}^+_\Delta| + |\mathcal{E}^-_\Delta| -
 2\left[v_F({\bf t} \cdot {\bf p})+ \sqrt{ v_F^2|{\bf\tilde{p}}| ^2 +
     \Delta_0^2 }\right]  \right\}.  \nonumber \\
\label{veffsup}
\end{eqnarray}
\end{widetext}
  Performing the momentum integrals in Eq.~(\ref{veffsup}), one finds
\begin{eqnarray}\label{potsup}
\lefteqn{\Omega^{ren}(0,\Delta_0, \mu ) = \frac{\Delta_0^2}{2 g_2 v_F}
  + \frac{(\mu^2 + \Delta_0^2)^{3/2}}{3 \pi v_F^2\xi_x \xi_y} }
\nonumber \\ && - \frac{\mu^2 \sqrt{\mu^2 + \Delta_0^2}}{2 \pi
  v_F^2\xi_x \xi_y} - \frac{\mu \Delta_0^2}{2 \pi v_F^2 \xi_x
  \xi_y}\ln \left(\frac{\mu+ \sqrt{\mu^2 + \Delta_0^2}}{\Delta_0} \right)
\nonumber\\ && +\frac{1}{2 \pi \xi_x \xi_y v_F^2} I(\Delta_0,\mu),
\label{omegasup}
\end{eqnarray}
where the function $I(\Delta_0,\mu)$ is derived explicitly in the
Appendix~\ref{appB} and given by Eq.~\eqref{Idelta}. The effective
thermodynamic potential given by Eq.~(\ref{omegasup}) is shown in
{}Fig.~\ref{fig1}(a) for $g_2<0$, while for $g_2>0$ it is shown in
{}Fig.~\ref{fig1}(b) for  $g_2>0$, where we have considered some
representative values of the effective tilt parameter and for the
chemical potential.  The emergence of a superconducting
gap $\Delta \neq 0$ due to the  combined effect of the tilt and
chemical potential is noted.  Let us analyze in more details the contribution
of the tilt parameter for the superconducting gap. New features
generated by the tilt of the Dirac cone will influence the
superconducting gap for both the $g_2>0$ and $g_2<0$ scenarios and are
explained below.

\begin{center}
\begin{figure}[!htb]
\subfigure[$g_2<0$]{\includegraphics[width=7.5cm]{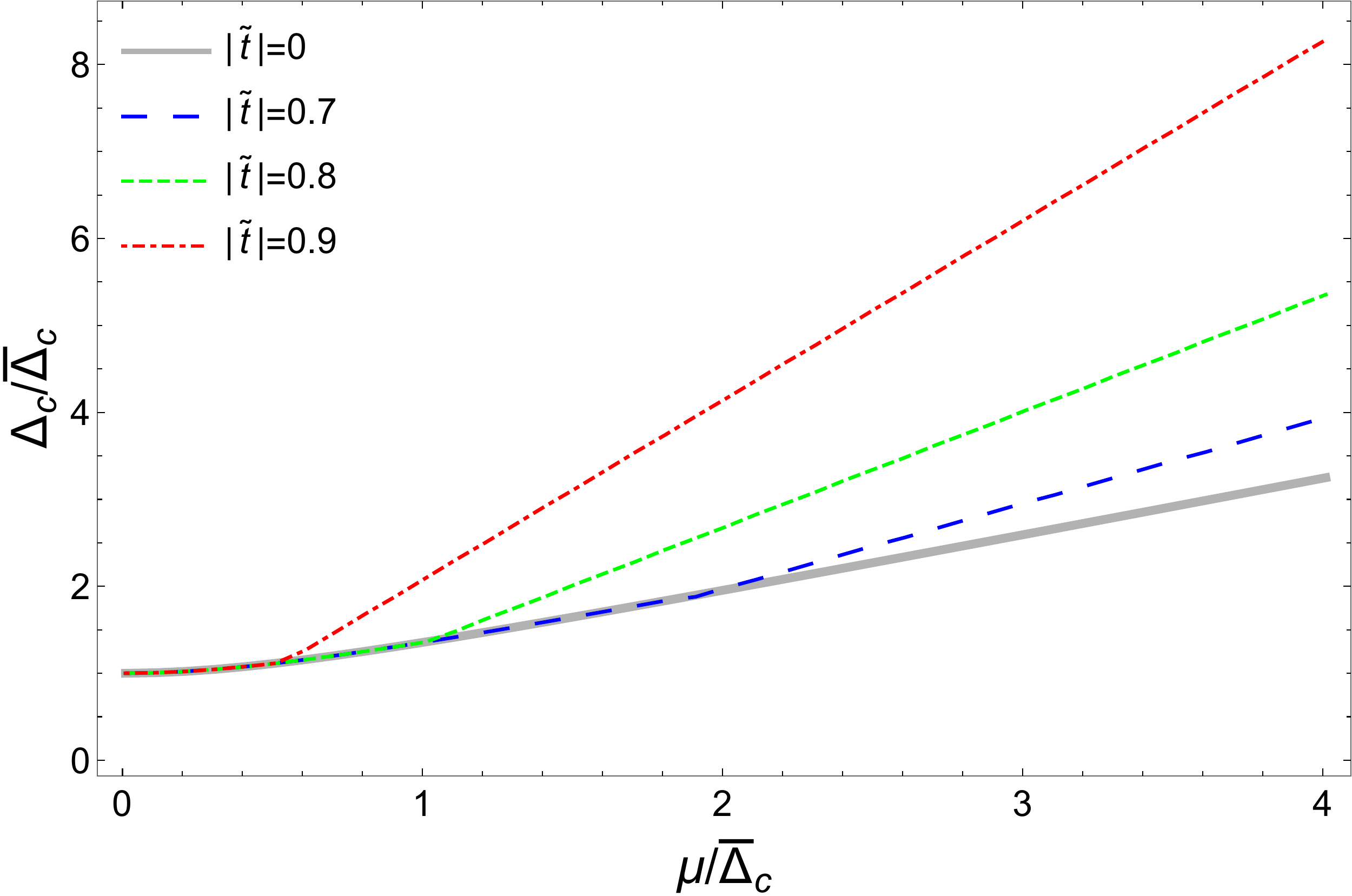}}
\subfigure[$g_2>0$]{\includegraphics[width=7.5cm]{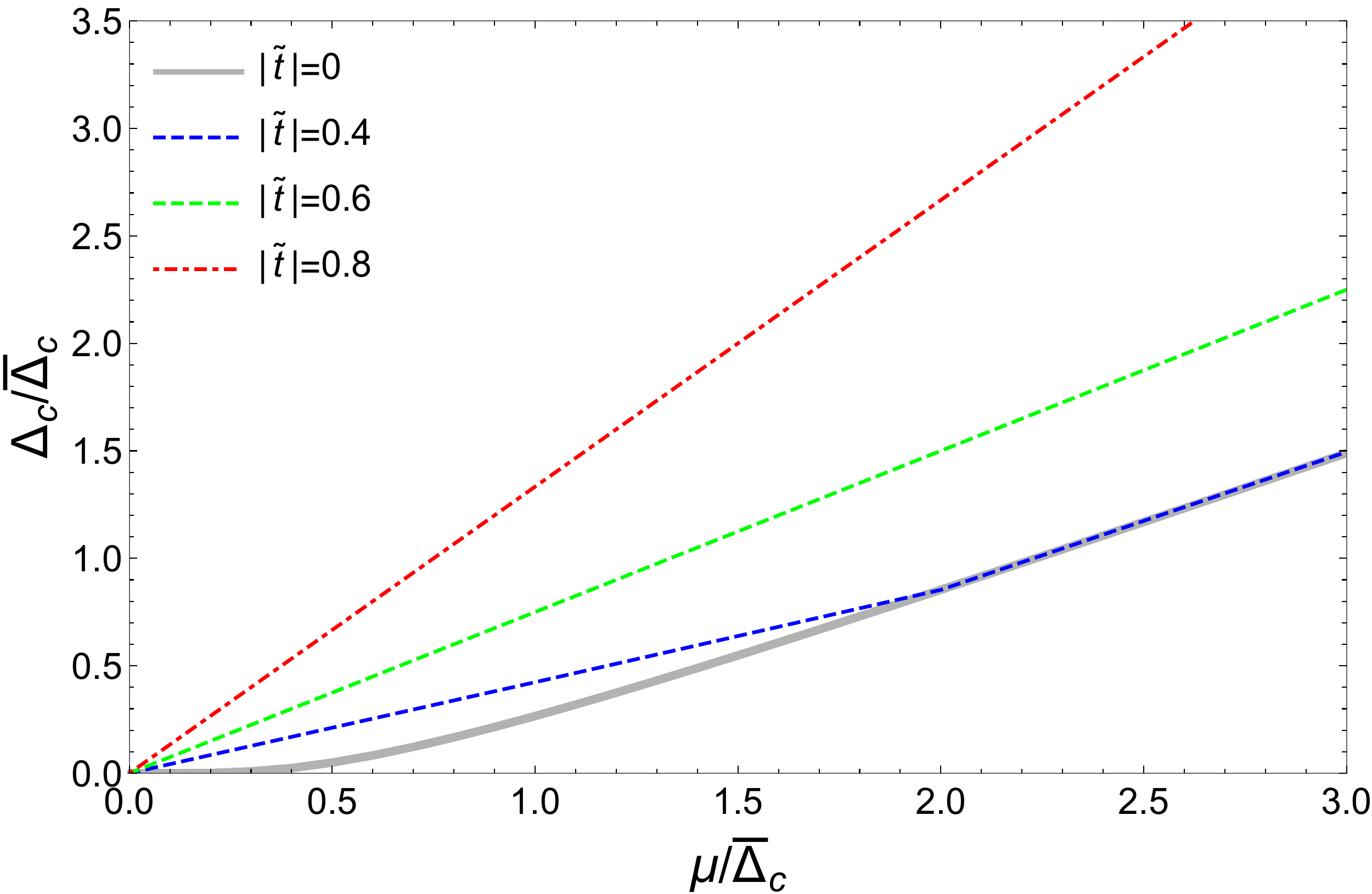}}
\caption{Superconducting gap for (a) $g_2<0$  and (b) $g_2>0$
induced by the chemical potential and the tilt parameter
  in units of $\bar{\Delta}_c $ for some representative values of
  $|{\bf \tilde{t}}|$.}
\label{fig2}
\end{figure}
\end{center}

{}From the effective thermodynamic potential one derives the gap
equation,
\begin{eqnarray}
&&{\rm sign} (g_2) + \sqrt{x^2 + y^2} - y \ln \left( \frac{y +
      \sqrt{x^2 + y^2}}{x}\right)  \nonumber \\ && + \frac{1}{2 x}
  \frac{\partial I(x,y)}{\partial x} = 0,
\label{gapsup}
\end{eqnarray}
where $x = \Delta_0/\bar{\Delta}_c$ , $y =\mu/\bar{\Delta}_c$.   The
superconducting gap that is induced by the chemical potential and the
tilt parameter is shown in {}Figs.~\ref{fig2}(a) and \ref{fig2}(b),
for the cases of $g_2<0$  and for $g_2>0$, respectively.

In {}Fig.~\ref{fig2}, the numerical results for the superconducting
gap are shown as a function of the chemical potential and some
representative values for the effective tilt parameter $|\tilde{\bf
  t}|$. 

\begin{figure}[!htb]
\centering \includegraphics[width=7.5cm]{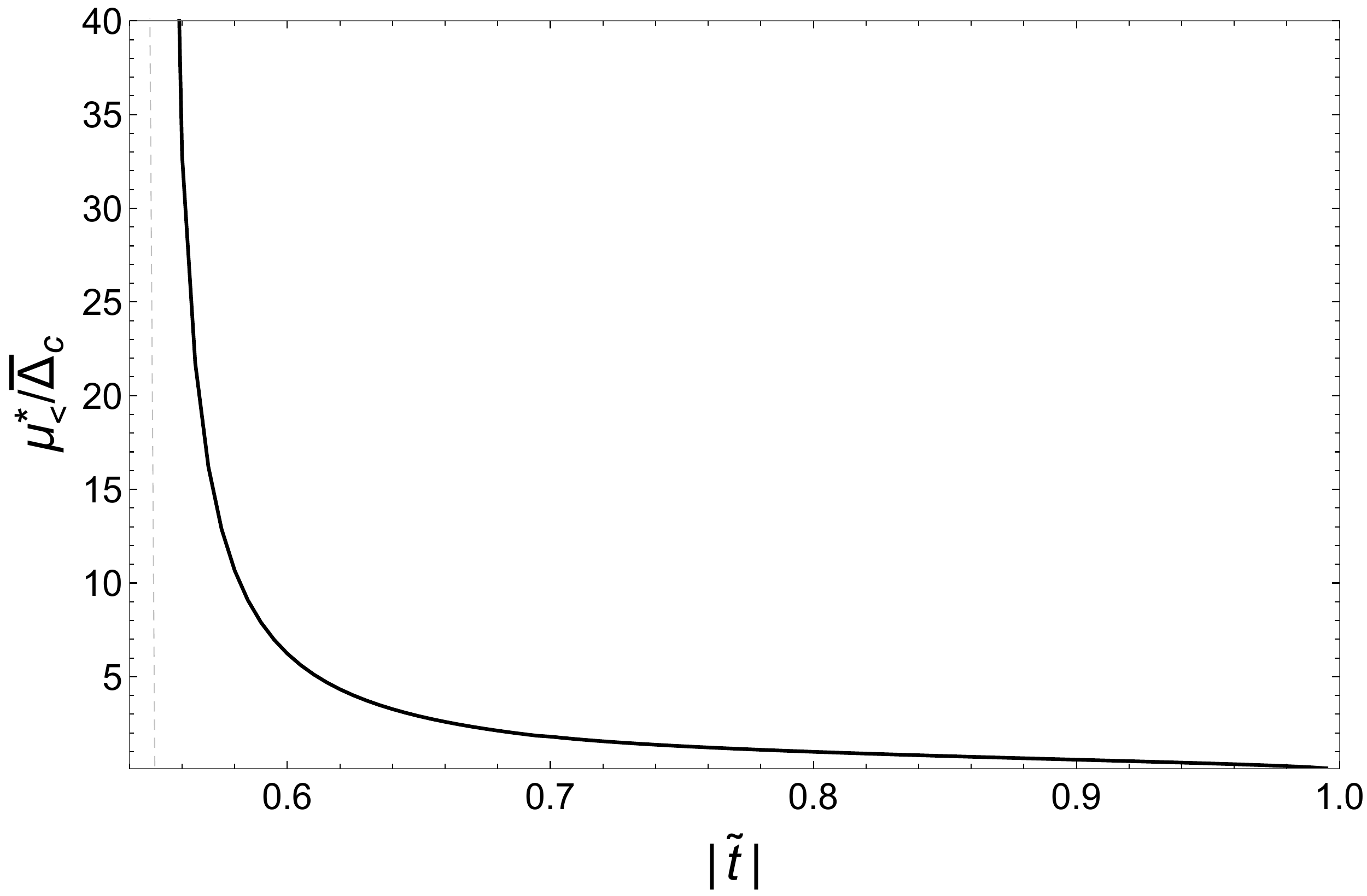} 
 \caption{The normalized chemical potential $\mu^*_{<}$ as a function
   of $|{\bf \tilde{t}}|$.  The thin vertical dashed line represents
   the threshold value $|{\bf \tilde{t}}| = \tilde{t}^*$.  }
 \label{fig3}
 \end{figure} 

In the case $g_2<0$, which is shown in {}Fig.~\ref{fig2}(a),  one can
see that the tilt increases $\Delta_c$ for a given $\mu >
\mu^*_{<}$. In the particular value $\mu^*_{<} = \mu^*_{<}(|{\bf
  \tilde{t}}|)$ is where the tilt parameter starts to contribute to
the superconducting gap. The behavior of  $\mu^*_{<}$ as a function of
$|{\bf\tilde{t}}|$ is shown in {}Fig.~\ref{fig3}. We find that there
is a threshold value for the effective tilt parameter, $\tilde{t}^*$,
such that when $|{\bf \tilde{t}}| < \tilde{t}^*$, the superconducting
gap is given by $\Delta_0 =\Delta_{{\bf t}=0}$ for any $\mu$. However,
for values of $|{\bf \tilde{t}}| > \tilde{t}^*$ and when  $\mu >
\mu^*_{<}$, the superconducting gap is given by $\Delta_0 =
\Delta_{{\bf t}}$. On the other hand, in the case of $\mu <
\mu^*_{<}$, the superconducting gap is given instead again by
$\Delta_0 = \Delta_{{\bf t}=0}$.  Let us obtain an explicit estimation
for this particular value for the effective tilt parameter,
$\tilde{t}^*$.  {}For $\mu>\mu^*_{<}$, the superconducting gap takes
the exact form $ \Delta_c (\mu>\mu^*_{<}) = \Delta_{ \bf t}$, where
\begin{equation}\label{sgapt}
\Delta_{{\bf t}}= \frac{|{\bf \tilde{t}}|\mu}{\sqrt{1-|{\bf
      \tilde{t}}|^2}}.
\end{equation}
Moreover, in order to extract the asymptotic behavior of the
superconducting gap shown in {}Fig.~\ref{fig2}, one first notes that
in the gap equation for the non-tilted case ($\tilde{ \bf t}=0$), the
last term in Eq.~(\ref{gapsup}) vanishes. Hence, for $\tilde{ \bf
  t}=0$,
\begin{eqnarray}\nonumber\label{app31}
&&{\rm sign} (g_2) + \sqrt{x^2 + y^2} - y \ln \left( \frac{y +
      \sqrt{x^2 + y^2}}{x}\right)  = 0.
\end{eqnarray}
Now, it is reasonable to assume that in the large $y$ limit the
normalized gap solution $x$ becomes a linear function of the
normalized chemical potential $y$, i.e., $x=\lambda y + c$, with $c$ a
constant. Hence, considering the asymptotic limit $y,x\ll c$  and
multiplying Eq.~\eqref{app31} by $1/y$, one obtain that $\lambda$
satisfies
\begin{equation}
\sqrt{\lambda ^2+1}-\ln \left(\frac{\sqrt{\lambda ^2+1}+1}{\lambda
}\right)\approx 0.
\end{equation}
The above equation has one positive solution given by $\lambda \simeq
0.66$. The threshold value $\tilde{t}^*$ for which the effective tilt
parameter begins to drive the superconducting gap is determined when
the superconducting gap, given by Eq.~\eqref{sgapt}, becomes parallel
to the asymptotic linear behavior of the tilt-less gap equation, i.e.,
we must have $\Delta_{\bf t} = \lambda \mu $. This leads to the
relation 
\begin{equation}
\frac{|\tilde{ \bf t}|}{\sqrt{1-|\tilde{ \bf t}|^2}} = \lambda~~.
\label{lambda}
\end{equation}
The solution of the above equation gives us the result $\tilde{t}^*
\simeq 0.55$ when using the solution for $\lambda$ obtained from
Eq.~(\ref{lambda}). This result agrees quite well with the numerical
results expected from {}Figs.~\ref{fig2} and \ref{fig3}. 

\begin{figure}[!htb]
\centering \includegraphics[width=7.5cm]{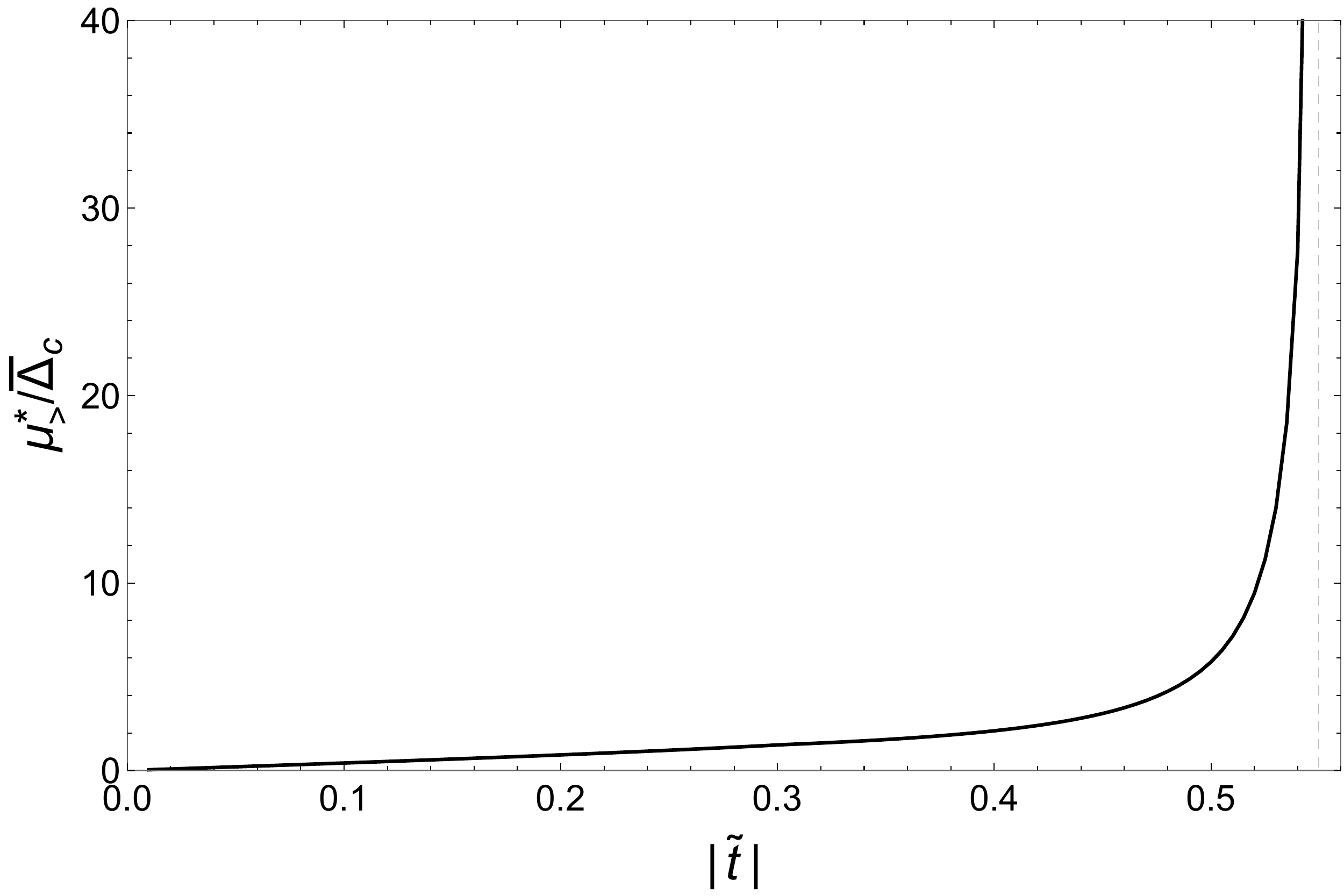} 
 \caption{Plot of the normalized chemical potential $\mu^*_{>}$. The
   thin vertical dashed line represents the threshold value $|{\bf
     \tilde{t}}| = \tilde{t}^*$. }
 \label{fig4}
 \end{figure}

When $g_2<0$, for any value of the tilt parameter $|{\bf \tilde{t}}|<
\tilde{t}^*$ the effect of the tilt parameter in the superconducting
gap vanishes for any $\mu$, and the superconducting gap of the system
obeys the solid gray curve shown in {}Fig.~\ref{fig2}(a).  We can also
analyze the situation for the case of $g_2>0$. Analyzing now the case
for $g_2>0$, we are able to uncover another structure for the
superconducting gap. As can be seen in {}Fig.~\ref{fig2}(b), in this
case we have two different situations.  When $|{\bf \tilde{t}}|<
\tilde{t}^*$, the tilt parameter only contributes for the chemical
potential up to the values $\mu_>^*$, $\mu< \mu^*_{>}$, e.g., as in the
case seen by the blue curve in {}Fig.~\ref{fig2}(b).  This particular
value $\mu^*_{>} = \mu^*_{>}(|{\bf \tilde{t}}|)$ sets a lower limit
where the tilt parameter stops contributing to the superconducting
gap. The behavior of $\mu^*_{>}$ is shown in {}Fig.~\ref{fig4}. When
$|{\bf \tilde{t}}|> \tilde{t}^*$,  the superconducting gap will be
exactly $ \Delta_c(|{\bf \tilde{t}}|>\tilde{t}^*) = \Delta_{{\bf t}}
$. As seen in {}Fig.~\ref{fig4}, we now have that when $g_2>0$,  for
values of  $\mu > \mu^*_{>}$ the superconducting gap is given by
$\Delta_0 = \Delta_{{\bf t}=0}$ and for $\mu < \mu^*_{>}$ the
superconducting gap is given by $\Delta_0 = \Delta_{{\bf t}}$.  {}For
$|{\bf \tilde{t}}| > \tilde{t}^*$ the superconducting gap is given by
$\Delta_0 =\Delta_{{\bf t}}$ for any $\mu$.  

{}Finally, we can explicitly compute the charge density, 
\begin{equation}
n = -N \frac{\partial \Omega^{ren}(0,\Delta_0,\mu)}{\partial \mu}
\Bigg{|}_{\Delta_0 = \langle \Delta_0 \rangle},
\end{equation}
which it can be expressed through an exact expression and given by
\begin{eqnarray}
n &=& \frac{N}{4 \pi v_F^2 \xi_x \xi_y}\left[ \mu\sqrt{\mu^2 +
    \Delta_0^2}  \right.  \nonumber \\ &+ & \left. \Delta^2_0 \ln
  \left(\frac{\mu + \sqrt{\mu^2 + \Delta_0^2}}{\Delta_0} \right)
  \right] \Bigr|_{\Delta_0 = \langle \Delta_0 \rangle}  \nonumber
\\ &+ &N \frac{\partial I(\Delta_0,\mu)}{\partial \mu}
\Bigr|_{\Delta_0 = \langle \Delta_0 \rangle},
\end{eqnarray}
where $\langle \Delta_0 \rangle$ is the solution of
Eq.~\eqref{gapsup}, which can be found numerically for both the $g_2>0$
and $g_2<0$ cases. {}From the inequality ${\rm Re} \sqrt{ |{\bf
    \tilde{t}}|^2\mu^2 -(1-|{\bf \tilde{t}}|^2) \Delta_0^2}\neq 0$,
one finds that the contribution for the charge density from the
function $I$ is non-null only for $\langle \Delta_0 \rangle <
\Delta_{\bf t}$. Thus, based on {}Fig.~\ref{fig2}(a), this
contribution is non-null only for $\mu< \mu^*_<$. In the case where
$g_2>0$, on the other hand, from {}Fig.~\ref{fig2}(b),  the density
will receive extra contributions only for $\mu>\mu^*_>$. 
 
\section{Phase structure for $\mu \neq 0$}
\label{sec5}
 
Previous
works~\cite{Klimenko:2012tk,Ebert:2016ygm,Zhukovsky:2017hzo,Klimenko:2012tk}
have shown that it is sufficient to analyze the chiral-superconducting
phase structure by comparing the vacuum properties in the $\sigma_0=0$
and $\Delta_0=0$ axes. Here we follow the same strategy. Through this
analysis of the local minimum in each axis, we can compare them and
find the global minimum which defines the real phase of the
system. {}For instance, as shown in the previous section, for fixed
$g_1<0$, there is a chemical potential for coexistence,
$\mu_c(g_2)$. The value of  $\mu_c(g_2)$  defines the lower bound for
the chemical potential such that for $\mu> \mu_c(g_2)$ the system is
in the superconducting phase (phase III), for $\mu< \mu_c(g_2)$ the
system is in the chiral symmetry-breaking phase (phase II), and for
$\mu = \mu_c(g_2)$ both phases II and III coexists.   This coexistence
point $\mu=\mu_c(g_2)$ defines a first-order transition between
phases II and III.  In the case of $g_2 <0$, there is another
particular value for the chemical potential, $\mu_<^*$, as discussed
in the previous section, such that for $\mu> \mu_<^*$ the
superconducting phase stops to drive the system in favor of the chiral
phase. The opposite happens when $g_2 >0$, in which case there is now
a value for the chemical potential,  $\mu=\mu_>^*$ that becomes an
upper bound and, for  $\mu < \mu_>^*$, it is when the superconducting
phase stops to drive the system in favor of the chiral
phase. {}Finally, the chiral symmetry will be restored for $\mu>
\sqrt{1-|\bf \tilde{t}|^2} \bar{\sigma}_c$.  Let us now show the
different phase portraits that will display the above structure
relating the chemical potential with the superconducting coupling
constant $g_2$ of the system when assuming $g_1 < 0$, which is the
relevant situation for nontrivial chiral and superconducting gaps. 

\begin{center}
\begin{figure}[!htb]
\subfigure[]{\includegraphics[width=7.5cm]{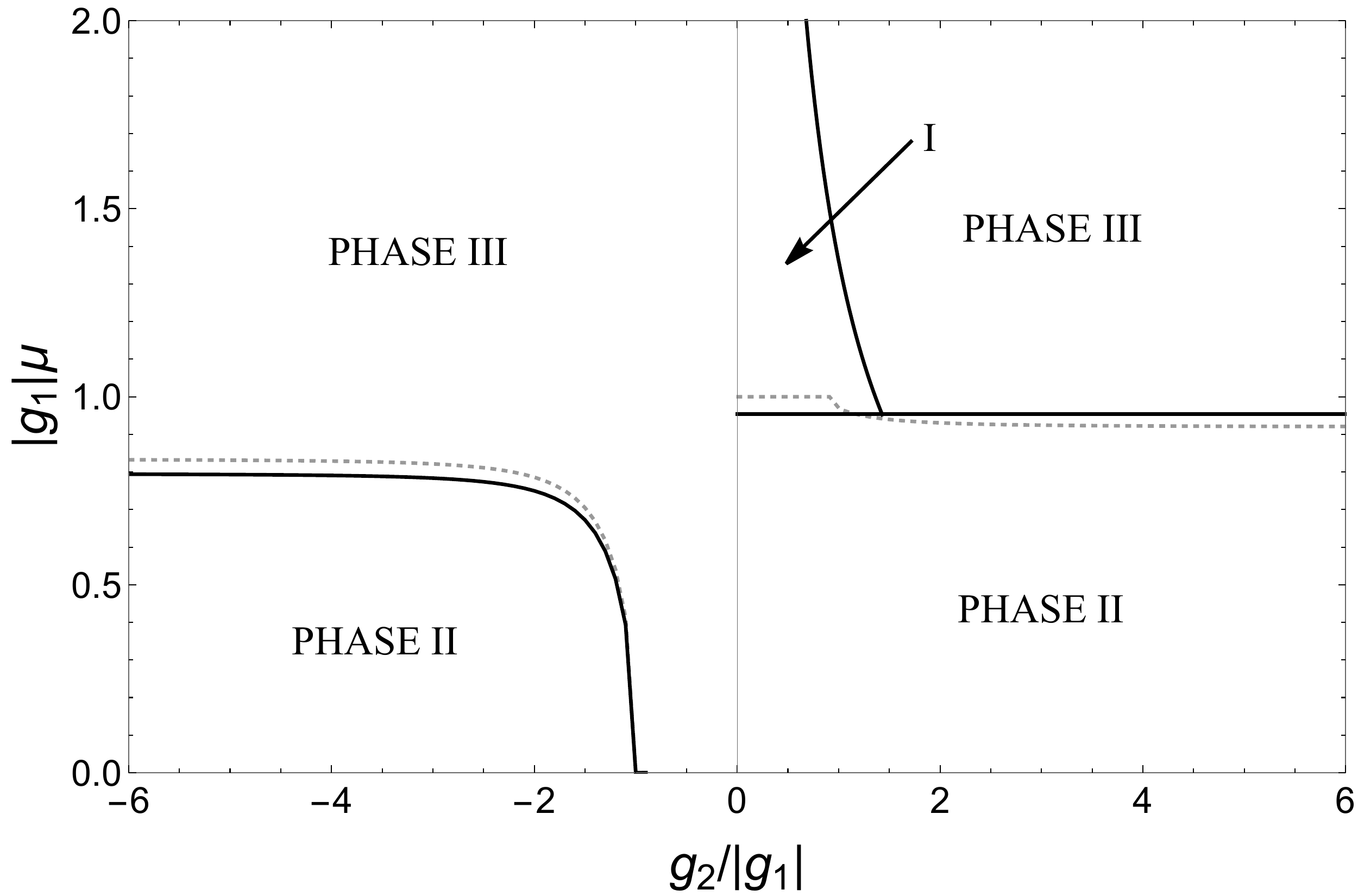}}
\subfigure[]{\includegraphics[width=7.5cm]{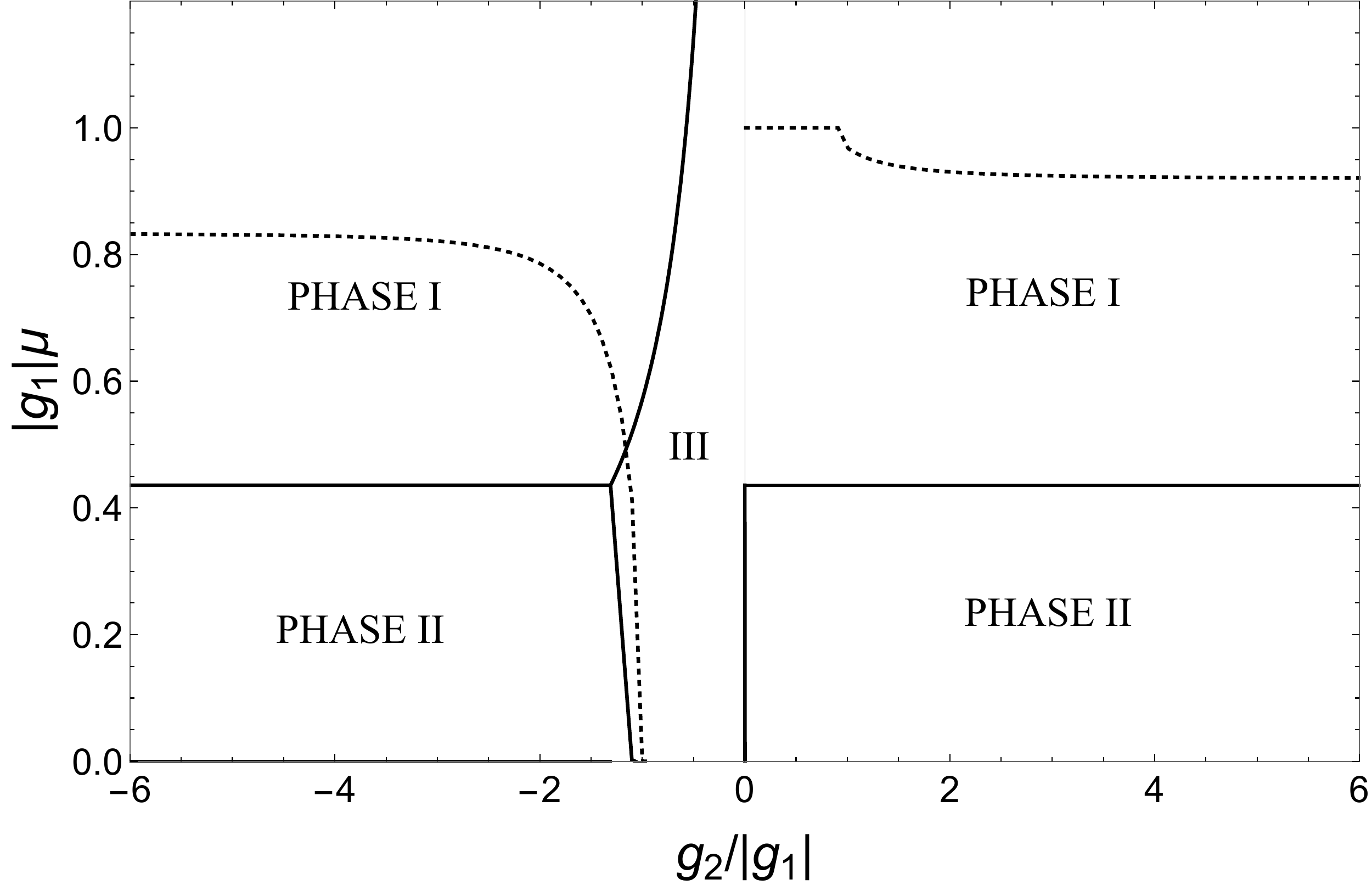}}
\caption{Phase portrait for the normalized chemical potential ($|g_1|\mu$)
  vs $g_2/|g_1|$ for $g_1<0$  and when (a) $|{\bf \tilde{t}}|=0.3 <
  \tilde{t}^*$ and (b) $|{\bf \tilde{t}}|=0.9>\tilde{t}^*$. 
The thin dashed lines represents the value for the
  chemical potential of coexistence in the nontilted case.
Phases I, II and III represent the metallic phase (with $\sigma_0 = \Delta_0 = 0$), 
the insulating phase (with  $\sigma_0  \neq 0$ and  $\Delta_0 =0$) and the 
superconducting phase (with  $\sigma_0=0$ and $\Delta_0  \neq 0$), respectively.}
\label{fig5}
\end{figure}
\end{center}

{}For illustration, in {}Fig.~\ref{fig5}(a) we show the phase portrait
when  $|{\bf \tilde{t}}|=0.3 < \tilde{t}^*$ and in the region ranging
from negative to positive values for $g_2$, while in
{}Fig.~\ref{fig5}(b) the phase portrait is shown for the case $|{\bf
  \tilde{t}}|=0.9 > \tilde{t}^*$.  {}For reference, in both
{}Figs.~\ref{fig5}(a) and \ref{fig5}(b), the phase
portrait in the nontilted case, $|{\bf \tilde{t}}|=0$, is shown
by the light gray dashed line, which matches the result previously
obtained in Ref.~\cite{Klimenko:2012tk}. Note that in the non-tilted
case, $|{\bf \tilde{t}}|=0$, the lines of coexistence separates the
phase II, which lies below the dashed line, from the phase III, which
lies above it. There is no phase I (where the chiral and
superconducting phases are absent) in this case.  Looking at the
region where $g_2<0$ in the case $|{\bf \tilde{t}}|< \tilde{t}^*$
shown in {}Fig.~\ref{fig5}(a), it is apparent that the presence of the
effective tilt parameter $|{\bf \tilde{t}}|$ does not qualitatively change
the phase portrait with respect to that of the
non-tilted case.  The structure of the phase transition can be
summarized as a first-order phase transition between the insulating
phase and the superconducting phase for a given $\mu_c(g_2)$
represented by the black line. In this case, the superconducting phase
is present for $\mu >\mu_c(g_2)$ in the same manner as in the
non-tilted situation.  Looking now at the region where $g_2>0$ for the
case $|{\bf \tilde{t}}|< \tilde{t}^*$ also shown in
{}Fig.~\ref{fig5}(a), one can notice that the superconductivity
induced by the chemical potential still exists, but in a smaller area
when compared to the non-tilted case (dashed line).  We recall that
from the results shown in the previous section, for $\mu < \mu^*_{>}$
and when $\mu> \sqrt{1-|{\bf \tilde{t}}|^2} \bar{\sigma}_c$, phase
I takes place. Thus, in this case, one finds a  point of coexistence,
$(\mu^{t},g_2^{t})$, which separates phases I-III, which is
given by
\begin{equation}\label{tric1}
(\mu^{t},g_2^{t})|_{g_2> 0,\, |{\bf \tilde{t}}|< \tilde{t}^* }=\left(
  \frac{\sqrt{1-|{\bf \tilde{t}}|^2}}{|g_1|},\frac{\sqrt{1-|{\bf
        \tilde{t}}|^2}}{\mu_{>}^*} \right)~.
\end{equation}
The presence of the coexistence point as a consequence of the tilt of
the Dirac cone is one of our main results, showing a quite different
behavior when compared to the results in the nontilted
case~\cite{Klimenko:2012tk}. 

Going further, looking at the case for $g_2<0$ and $|{\bf \tilde{t}}|>
\tilde{t}^*$, which is shown in {}Fig.~\ref{fig5}(b), one notices a
much stronger change in the phase portrait as compared to the region
with $g_2<0$ shown in {}Fig.~\ref{fig5}(a). The presence of the tilt
effectively causes the superconducting gap to stop to drive the system
for $\mu>\mu^*_{<}$ and phase I now takes place for $\mu>
\sqrt{1-|{\bf \tilde{t}}|^2}\bar{\sigma}_c$. The phase portrait in
this case displays a much restricted area for the superconducting
phase. The superconducting phase occurs only for values of
$|g_2|/|g_1| \lesssim 1$. In this case, one coexistence point also
appears and it is found to be given by
\begin{equation}\label{tric3}
(\mu^{t},g_2^{t})|_{g_2<0, \,|{\bf \tilde{t}}|> \tilde{t}^*}=\left(
  \frac{\sqrt{1-|{\bf \tilde{t}}|^2}}{|g_1|},-\frac{\sqrt{1-|{\bf
        \tilde{t}}|^2}}{\mu_{<}^*} \right).
\end{equation}
{}Finally, looking at the region where $g_2>0$ shown in
{}Fig.~\ref{fig5}(b), the induction of a superconducting phase due to
the chemical potential is ruled out for any value of the chemical
potential and the phase transition occurs between phases I and
II. Through the increase of the chemical potential and in the presence
of a tilt satisfying  $|{\bf \tilde{t}}|> \tilde{t}^*$, both effects
work in favor of the chiral symmetric phase. This can be seen by the
enlarged region for phase I shown in {}Fig.~\ref{fig5}(b) when
compared to the nontilted case. This is our other main result that is
extracted from the phase portrait. It shows once more the effect of
the tilt on hindering the formation of gaps in the system and, in this
case, the formation of an induced gap due to the presence of the
chemical potential. The role of the threshold value for the effective
tilt parameter $\tilde{t}^*$ becomes quite evident when contrasting
the two panels in {}Fig.~\ref{fig5}.

\section{Concluding remarks}
\label{conclusions}

In this paper, we have investigated the phase diagram of the Weyl
fermion system with four-fermion interactions that introduce the
effects of both chiral and superconducting gaps. {}Furthermore, we
have focused on the effect of the tilt factor of the Dirac cone. As
one of our main results, it is the demonstration, both analytically and
numerically, of the presence of a threshold value for the effective
tilt parameter $ \tilde{t}^*$ beyond which the value of the tilting of the
Dirac cone strongly affects the superconducting gap. More
specifically, one explicitly finds that $\tilde{t}^* \simeq 0.55$. The
stability of the superconducting phase is also found to be much
different, whether the tilting factor is lower or higher than $
\tilde{t}^*$.  At this value for the effective tilt parameter the
system behaves completely differently under the formation of the
chiral and superconducting gaps when compared to the nontilted
case. In the case where $|{\bf  \tilde{t}}|< \tilde{t}^*$, the
superconducting phase persists for a negative superconducting coupling
constant, which is responsible for the attractive interaction in the
Cooper channel. A first-order phase transition occurs for a chemical
potential for coexistence, as seen by the black curve in
{}Fig.~\ref{fig5}(a). This feature is similar to the results for
graphene and other two-dimensional
materials~\cite{Klimenko:2012tk}. One also sees that for $g_2>0$ the
induction of a superconducting gap due to the presence of a chemical
potential exists. This induction, however, only happens for stronger
values of the coupling constant $g_2$, since the metallic phase
appears for small values of the superconducting coupling constant. Due
to the presence of a metallic phase, we were able to find the
expression for the point of coexistence, which is given by
Eq.~\eqref{tric1}. 

While for values of $|{\bf  \tilde{t}}|< \tilde{t}^*$ the changes to
the phase portrait seen in {}Fig.~\ref{fig5}(a) are of a qualitative
nature, when the effective tilt exceeds the threshold value, the
changes now become quantitative.  When $|{\bf \tilde{t}}|$  exceeds
the value $\tilde{t}^*$, the superconducting phase now becomes
restricted to a smaller area in the phase portrait. Indeed, in this
case, the superconducting phase occurs only for regions with small and
negative superconducting coupling constant. Through a first-order
phase transition, the metallic phase takes place for a sufficient
large superconducting coupling constant and chemical potential. One
also finds the analytic expression for the point of coexistence in
this case, which is given by  Eq.~\eqref{tric3}. {}Finally, one shows
that our qualitative analysis point to the fact that for $g_2>0$, the
superconducting gap induced by the chemical potential is ruled out and
a first-order phase transition occurs between phases I and II at the
chemical potential for coexistence, which is represented by the black
curve in {}Fig.~\ref{fig5}(b).   The presence and role of the
threshold value for the effective tilt parameter represent one of the
main important results shown in this paper.  

 We can try to explore the consequences of the results we have
 obtained for some known planar systems and which have been currently
 studied in laboratory experiments.  {}For example, using the
 experimental data obtained from the two-dimensional (2D) organic conductor
 $\alpha-\text{(BEDT-TTF)}_2 I_3$~\cite{ex1,ex2}, the estimated
 effective tilt parameter is found to be $|{\bf \tilde{t} }| \simeq
 0.76$ (see, e.g., Ref.~\cite{Gomes:2021nem}).  This case occurs in the
 situation where $|{\bf \tilde{t}}|> \tilde{t}^* \simeq 0.55$ and
 which we have discovered in this paper.  {}From our results, this
 implies that the inducing of a superconducting gap should be absent
 in this material. It would be interesting to probe this prediction
 using this type of material in the laboratory. By also accounting for
 the results obtained from the analysis of Ref.~\cite{Gomes:2021nem},
 we can also conclude that this same system should exhibit a metallic
 phase, which would become very strong under doping.  On the other
 hand, we can also compare with the predictions that our results would
 imply for the case of quinoid-type graphene under uniaxial
 strain~\cite{goerbig1}. In this case,  the estimated values for the
 effective tilt parameter are such that $|{\bf \tilde{t}}| \lesssim
 0.06$ for moderate deformations. {}From our results, we can conclude
 that for this material the properties of the superconducting gap
 should be similar to the graphene case, which includes the induction
 of a superconducting gap by the chemical potential.  To the authors'
 best knowledge, we are not aware of other materials where the value
 of the tilt parameter has been provided, at least as far as
 two-dimensional materials are concerned. We are hopeful that as new
 two-dimensional materials are experimentally probed and fabricated,
 new data from those experiments will help to shed light on the
 results we have presented here.

The study of possible two-dimensional fermionic systems where our
results can be of interest can be exploited in several
directions. {}First, since the evaluation of $\tilde{t}^*$ is based
on the large-$N$ limit of the effective thermodynamical potential, it
is possible that this result receives quantum corrections beyond the
large-$N$ approximation. This can be an interesting extension of the
present work. {}Going further, the presence of an anomalous Hall
effect~\cite{Gomes:2021nem,AHE,AHE2} in the 2D Weyl semimetal
indicates the possibility that the tilt of the Dirac cones could
modify the superconducting gap under the presence of an external
magnetic field.  Moreover, since the tilt of the Dirac cone introduces
a special direction in the system, the analysis of the $p$-wave
superconducting gap properties in this context could bring new
features. This can be another problem of interest that can be a target
of further investigation. These problems are possible lines of study
that our results motivate and we hope to address them in the future. 

\begin{acknowledgments}

Y.M.P.G. is supported by a postdoctoral grant from  {}Funda\c{c}\~ao
Carlos Chagas Filho de Amparo \`a Pesquisa do Estado do Rio de Janeiro
(FAPERJ).  R.O.R. acknowledges financial support of the
Coordena\c{c}\~ao de Aperfei\c{c}oamento de Pessoal de N\'{\i}vel
Superior (CAPES) - Finance Code 001 and by research grants from
Conselho Nacional de Desenvolvimento Cient\'{\i}fico e Tecnol\'ogico
(CNPq), Grant No. 307286/2021-5, and from Funda\c{c}\~ao Carlos Chagas
Filho de Amparo \`a Pesquisa do Estado do Rio de Janeiro (FAPERJ),
Grant No. E-26/201.150/2021. 

\end{acknowledgments}

\begin{appendix}

\section{ Performing the path integral over the fermion in
  Eq.~(\ref{effaction})}
\label{appA}

Here we show some of the details of the path integral over the
fermions in Eq.~(\ref{effaction}), which leads to the effective
thermodynamic potential.  Adopting the procedure described in
Ref.~~\cite{Klimenko:2012tk}, we assume two anti-commuting four-component 
Dirac spinor fields $q(x)$ and $\bar{q}(x)$. Then,
Eq.~\eqref{effaction} can be rewritten as
\begin{eqnarray}
I &=& \int Dq D\bar{q}~ e^{ i \int d^3x \left[ \bar{q} \mathcal{O} q -
    \frac{\Delta}{2} q^T C q - \frac{\Delta^*}{2} \bar{q} C \bar{q}^T
    \right] },
\end{eqnarray}
where $\mathcal{O} = i M^{\mu \nu} \gamma_\mu \partial_\nu + \mu
\gamma^0- \sigma$ and $C = i \gamma^2$ is the charge conjugation
matrix. Using the Gaussian path integral identities
\begin{eqnarray}
&&\int Dp e^{ i \int d^3x \Big[ -\frac{1}{2} p^T A p + \eta^T p \Big]
  } \nonumber \\ &=& (\det A)^{\frac{1}{2}} e^{ -\frac{i}{2} \int d^3x
    \eta^T A^{-1} \eta }~,
\end{eqnarray}
and 
\begin{eqnarray}
&&\int D\bar{p} e^{ i \int d^3x \Big[ -\frac{1}{2} \bar{p} A \bar{p}^T
      + \eta \bar{p}^T \Big] } \nonumber\\ &=& (\det A)^{\frac{1}{2}}
  e^{ -\frac{i}{2} \int d^3x \bar{\eta} A^{-1} \bar{\eta}^T },
\end{eqnarray}
and by also considering $A = \Delta C$, $\bar{q} \mathcal{O} =
\eta^T$, $\mathcal{O}^T \bar{q}^T = \eta$, one finds, after
integrating over $q$ and $\bar{q}$, the result
\begin{eqnarray}
I &=& \int Dq D\bar{q}~ e^{ i \int d^3x \Bigg[ \bar{q} \mathcal{O} q -
    \frac{\Delta}{2} q^T C q - \frac{\Delta^*}{2} \bar{q} C \bar{q}^T
    \Bigg] } \nonumber \\ &=& (\det \Delta C)^{\frac{1}{2}} \int
D\bar{q} e^{ \frac{i}{2} \int d^3x \Bigg\{ \bar{q} \left[ \Delta^* C +
    \mathcal{O} (\Delta C)^{-1} \mathcal{O}^T\right]\bar{q}^T  \Bigg\}
}  \nonumber\\ &=& (\det \Delta C)^{\frac{1}{2}} \left[\det(\Delta^* C
  + \mathcal{O} (\Delta C)^{-1} \mathcal{O}^T)\right]^{\frac{1}{2}}
\nonumber\\ &=& \left[\det \left( \Delta^2 + \mathcal{O} C^{-1}
  \mathcal{O}^T C \right) \right]^{\frac{1}{2}}, \nonumber \\ 
\end{eqnarray} 
where we have assumed $\Delta = \Delta^*$ in the last step (we are not
interested in the phase of the superconducting order parameter, but
solely on its absolute (modulus) value). Using the relations $C^{-1}
\gamma_\mu^T C = - \gamma_\mu$ and $\partial_\mu^T = - \partial_\mu$
one finds that
\begin{equation}
 I = [\det(-\Delta^2 + \mathcal{O}_+ \mathcal{O}_-)]^{1/2} = (\det
 B)^{\frac{1}{2}},
\end{equation}
with $\mathcal{O}_\pm = i M^{\mu \nu} \gamma_\mu \partial_\nu \pm \mu
\gamma^0- \sigma$. {}Finally, using the identity $\det B = \exp({\rm
  Tr } \ln B)$ one finds
\begin{equation}
 \ln I = \frac{1}{2} {\rm tr} ( \ln B )= \int d^3x  \sum_{i=1}^2 \int
 \frac{d^3p}{(2 \pi)^3} \ln \lambda_i(p),
\end{equation}  
where 
\begin{eqnarray}
\lambda_{1,2} &=& \sigma^2 + \left[ p_0 - v_F ({\bf t} \cdot {\bf p} )
  \right]^2 - v_F^2{\bf \tilde{p}}^2 - \mu^2 - |\Delta|^2  \nonumber
\\ &\pm & 2 \sqrt{\sigma^2 \left\{ \left[p_0 - v_F({\bf t} \cdot {\bf
      p})\right]^2 - v_F^2{\bf \tilde{p}}^2 \right\} + v_F^2\mu^2{\bf
    \tilde{p}}^2 }, \nonumber \\
\label{eigenvalueslambda}
\end{eqnarray}
are the eigenvalues of $B$.

\section{ The effective thermodynamic potential}\label{appB}

In this section one shows some of the details for the derivation of
the effective thermodynamic potential.  {}From Eq.~\eqref{veffsup}, we
obtain
\begin{eqnarray}
\lefteqn{ \Omega^{ren}(0,\Delta_0, \mu ) = \frac{\Delta_0^2}{2 g_2
    v_F} + \frac{\Delta_0^3}{3 \pi v_F^2 \xi_x \xi_y} }
\nonumber\\  &&-   \int \frac{d^2p}{(2 \pi)^2}\left\{\frac{}{}
|\mathcal{E}^+_\Delta| + |\mathcal{E}^-_\Delta| \right.  \nonumber
\\ &&\left. - 2\left[v_F({\bf t} \cdot {\bf p})+ \sqrt{
    v_F^2|{\bf\tilde{p}}| ^2 + \Delta_0^2 }\right]  \right\},
\end{eqnarray}
where  $\mathcal{E}^\pm_\Delta =v_F({\bf t} \cdot {\bf p}) +\sqrt{
  (v_F|{\bf \tilde{p}}| \pm \mu)^2 + \Delta_0^2  }$. The effective
thermodynamic potential depends on momentum integrals of the form
\begin{eqnarray}
i_\pm&=& \int \frac{d^2p}{(2 \pi)^2}\left\{ |\mathcal{E}^\pm_\Delta|
-\left[ v_F({\bf t} \cdot {\bf p})+ \sqrt{ v_F^2|{\bf\tilde{p}}| ^2 +
    \Delta_0^2 } \right]  \right\}.  \nonumber \\
\end{eqnarray}
Then, $\Omega^{ren}(0,\Delta_0, \mu )$ can be written as
\begin{equation}
\Omega^{ren}(0,\Delta_0, \mu ) = \frac{\Delta_0^2}{2 g_2 v_F} +
\frac{\Delta_0^3}{3 \pi v_F^2 \xi_x \xi_y}- i_+-i_-. 
\end{equation}
It can now be shown that for $\mu>0$, $\mathcal{E}^+_\Delta > 0$ for
all $p>0$. Hence, after some algebraic steps, one finds
\begin{eqnarray}
i_+&=&  \frac{1}{2 \pi v_F^2 \xi_x \xi_y}\int_0^\infty dp p \left[
  \sqrt{ (p + \mu)^2 + \Delta_0^2  }- \sqrt{ p ^2 + \Delta_0^2
  }\right] .  \nonumber\\
\end{eqnarray}
{}For $\mathcal{E}^-_\Delta$, one has that $\mathcal{E}^-_\Delta>0$
only for $p<p_-$ and for $p>p_+$, where
\begin{equation}
p_\pm =\frac{1}{(1 - |{\bf \tilde{t}}|^2)} \Big[ \mu \pm {\rm Re}
  \sqrt{|{\bf \tilde{t}}|^2 \mu^2 - (1 - |{\bf
      \tilde{t}}|^2)\Delta_0^2} \Big],
\end{equation}
where ${\rm Re}$ means the real part. {}From the above expressions,
then, it follows that

\begin{eqnarray}
&&\int \frac{d^2p}{(2 \pi)^2}|\mathcal{E}^-_\Delta| = \frac{1}{2 \pi
    v_F^2 \xi_x \xi_y} \nonumber \\  && \times \int_0^\infty dp p
  \int_{0}^{2 \pi} \frac{d \theta}{(2 \pi)} \left[ \left| |{\bf
      \tilde{t}}| p \cos \theta  +\sqrt{ (p-\mu)^2 + \Delta_0^2  }
    \right| \right] \nonumber \\ &&=\frac{1}{2 \pi v_F^2 \xi_x \xi_y}
  \int_0^\infty dp p \sqrt{ (p-\mu)^2 + \Delta_0^2  } \nonumber
  \\ &&\times \Theta\left[ \sqrt{ (p-\mu)^2 + \Delta_0^2  }-|{\bf
      \tilde{t}}| p \right],
\end{eqnarray}
where we have used the identity $\int_0^{2 \pi} \frac{d \theta}{2 \pi}
| a \cos \theta + b| =  b \Theta(b-a)$, for $a>0$ and $b>0$ in the
last step and $\Theta(x)$ is the Heaviside function. {}Finally, the
inequality $\sqrt{ (p-\mu)^2 + \Delta_0^2  }-|{\bf \tilde{t}}| p>0$ is
respected when $p<p_-$ and $p>p_+$. In particular, one notes that in
the limit $|{\bf \tilde{t}}| \rightarrow 0 $, one finds $p_+ = p_- =
\mu$.  Therefore, using the fact that $\int_0^{p_-} +
\int_{p^+}^\infty = \int_0^\infty - \int_{p_-}^{p_+}$ it follows that
\begin{eqnarray}
i_- &=& \frac{1}{2 \pi v_F^2 \xi_x \xi_y}\int_0^\infty dp p \left[
  \sqrt{ (p - \mu)^2 + \Delta_0^2  }- \sqrt{ p^2 + \Delta_0^2 }\right]
\nonumber \\ &-& \frac{1}{2 \pi v_F^2 \xi_x \xi_y} \int_{p_-}^{p_+} dp
p \sqrt{ (p-\mu)^2 + \Delta_0^2  }  \nonumber \\ &=& \frac{1}{2 \pi
  v_F^2 \xi_x \xi_y}\Bigg(\int_0^\infty dp p \left[ \sqrt{ (p - \mu)^2
    + \Delta_0^2  }- \sqrt{ p ^2 + \Delta_0^2 }\right]
\nonumber\\ &-& I(\Delta_0, \mu) \Bigg),
\end{eqnarray}
where
\begin{eqnarray}
I(\Delta_0,\mu) &=& \int_{p_-}^{p_+} dp p \sqrt{ (p-\mu)^2 +
  \Delta_0^2  } \nonumber \\ &=& \left(2 \Delta_0 ^2-\mu ^2+2
p_+^2-\mu  p_+\right) \sqrt{\Delta_0 ^2+(p_+-\mu) ^2} \nonumber \\ &+&
3 \Delta_0 ^2 \mu  \tanh ^{-1}\left[\frac{p_+-\mu }{\sqrt{\Delta_0
      ^2+(p_+-\mu)^2}}\right] \nonumber \\ &-& \left(2 \Delta_0 ^2-\mu
^2+2 p_-^2-\mu  p_-\right) \sqrt{\Delta_0 ^2+(p_--\mu) ^2} \nonumber
\\ &-& 3 \Delta_0 ^2 \mu \tanh ^{-1}\left[\frac{p_--\mu
  }{\sqrt{\Delta_0 ^2+(p_--\mu)^2}}\right].
\label{Idelta}
\end{eqnarray}
In normalized units $x = \Delta_0/\bar{\Delta}_c$, $y = \mu/
\bar{\Delta}_c$,  $I(x,y)$ is given by
\begin{eqnarray}
I(x,y) &=&   \left(2 x^2-y^2+2 z_+^2-y  z_+\right) \sqrt{x^2+(z_+-y)
  ^2} \nonumber \\ &+& 3 x^2 y  \tanh ^{-1}\left[\frac{z_+-y
  }{\sqrt{x^2+(z_+-y)^2}}\right] \nonumber  \\ &- &  \left(2 x^2-y
^2+2 z_-^2-y z_-\right) \sqrt{x^2+(z_--y) ^2} \nonumber \\ &- & 3 x^2
y \tanh ^{-1}\left[\frac{z_--y }{\sqrt{x^2+(z_--y)^2}}\right],
\end{eqnarray}
with $z_\pm$ defined as
\begin{equation}
z_\pm =\frac{1}{(1 - |{\bf \tilde{t}}|^2)} \Big[ y \pm Re \sqrt{|{\bf
      \tilde{t}}|^2 y^2 - (1 - |{\bf \tilde{t}}|^2)x^2} \Big].
\end{equation}
 
{}Finally, after integration over the momentum $p$, one can write the
renormalized effective thermodynamic potential for the superconducting
phase (when $\sigma_0=0$) as
\begin{eqnarray}\label{app2pot}
\Omega^{ren}(0,\Delta_0, \mu ) &=& \frac{\Delta_0^2}{2 g_2 v_F} +
\frac{(\mu^2 + \Delta_0^2)^{3/2}}{3 \pi v_F^2\xi_x \xi_y} \nonumber
\\ &-& \frac{\mu^2 \sqrt{\mu^2 + \Delta_0^2}}{2 \pi v_F^2\xi_x \xi_y}
\nonumber \\ &-& \frac{\mu \Delta_0^2}{2 \pi v_F^2 \xi_x \xi_y}\ln
\left( \frac{\mu+ \sqrt{\mu^2 + \Delta_0^2}}{\Delta_0} \right)
\nonumber\\ &+&\frac{1}{2 \pi \xi_x \xi_y v_F^2} I(\Delta_0,\mu)
. \nonumber\\
\end{eqnarray}


\end{appendix}


\end{document}